# Reversible Data Hiding in Encrypted Images based on MSB Prediction and Huffman Coding

Youzhi Xiang[1], Zhaoxia Yin[1,*], Xinpeng Zhang[2]

[1] School of Computer Science and Technology, Anhui University
[2] School of Computer Science, Fudan University
* Corresponding email: yinzhaoxia@ahu.edu.cn

Contents

**Abstract**

    With the development of cloud storage and privacy protection, reversible data hiding in encrypted images (RDHEI) has attracted increasing attention as a technology that can embed additional data in the encryption domain. In general, an RDHEI method embeds secret data in an encrypted image while ensuring that the embedded data can be extracted error-free and the original image can be restored lossless. In this paper, A high-capacity RDHEI algorithm is proposed. At first, the Most Significant Bits (MSB) of each pixel was predicted adaptively and marked by Huffman coding in the original image. Then, the image was encrypted by a stream cipher method. At last, the vacated space can be used to embed additional data. Experimental results show that our method achieved higher embedding capacity while comparing with the state-of-the-art methods.

## I. Introduction

Reversible data hiding (RDH) is a technique to embed secret information into a cover image by slightly modifying pixel values. Existing RDH methods are mainly divided into three categories: lossless compression [1], histogram shifting [2] and difference expansion [3]. These methods are designed to ensure that secret information is not detected, and the secret data, as well as the original image, can be completely restored from the marked image. Due to this reversible feature, the RDH method can be applied in many fields, such as medical and military images. For an RDH method in the plaintext domain, rate distortion of an image is generally used to evaluate its performance, that is, to maximize the embedding rate while minimizing image distortion. Therefore, various RDH algorithms have been proposed to achieve better rate-distortion performance [4]-[8].

Recently, with the development of cloud computing and cloud storage, privacy protection has received widespread attention. The marked image uploaded to the cloud using a plaintext RDH method exposed the original content of the cover image, which is a result that the content owner does not want to see. Therefore, in order to solve this problem, many researchers show their interests in developing reversible data hiding methods in encrypted images (RDHEI) [9]-[19]. In these methods, there are three users: the content-owner, data-hider, and receiver. The content-owner encrypts the original image and send it to the data-hider. The data-hider embeds some secret information, e.g., the image source information or timestamp, into the encrypted image and cannot obtain the original image. On the receiving side, depending on its authority, the recipient can obtain the secret information or the original content of the image.

So far, many RDHEI methods have been proposed, and existing methods can be divided into two categories, namely vacating room after encryption (VRAE) [9]-[14] and reserving room before encryption (RRBE) [15]-[19]. In VRAE methods, the content-owner directly encrypts an original image and sends it to the data-hider, who then embeds some secret information by slightly modifying the encrypted image. Unlike the VRAE method, the RRBE method uses the spatial correlation of the original image to reserve room in the encrypted image before image encryption.

The idea of VRAE method was first proposed by *Puech et al.* [9] who encrypted the original image by using Advanced Encryption Standard, then they partitioned the encrypted image and embedded one bit of information in each block. Furthermore, data extraction and image restoration are achieved by analyzing the local standard deviation during decryption of the marked encrypted image. Different from [9], *Zhang* [10] uses a stream cipher to encrypt original image by exclusive-or (XOR) operation. Then the data-hider divides the encrypted image into blocks and each block is embedded with one bit of information by flipping three least significant bits (LSBs) of half the pixels in the block. At the receiving side, data extraction and image restoration are performed simultaneously, but the extracted data and the reconstructed image have an error rate. Based on this method, *Hong et al.* [11] proposed an improved method by exploiting the spatial correlation between neighboring blocks and using a side-match mechanism to obtain a higher embedding capacity with the lower error rate in image recovery. Note that in the methods of *Zhang* [10] and *Hong et al.* [11], the embedding capacity is related to its block size, data extraction and image restoration are inseparable.

After that, *Zhang* [12] designed a separable RDHEI method. First, a content-owner encrypts the original image with an encryption key. Then, the data-hider uses a data hiding key to compress the LSBs of pixels in the encrypted image to vacate the room for storing secret information. On the receiving side, data extraction and image recovery can be performed separately according to different keys. *Wu and Sun* [13] proposed two RDH methods in encrypted images, namely a joint method and a separable method, are introduced by adopting prediction error. Both methods encrypt the original image in the same way as *Zhang* [12]. The difference is that data extraction and image recovery in the joint method are performed simultaneously, while the second method is separable. Another separable RDHEI method was proposed by *Qian and Zhang* [14], which is inspired by the distributed source coding. After the original image is encrypted by the content-owner using an encryption key, the data-hider compresses some bits selected from the encrypted image using low-density parity-check codes to make room for the secret data. Of course, this method can also achieve separation of data extraction and image recovery.

By analyzing the existing VRAE methods, it is found that the embedding capacity of these methods is relatively low, and there may be subject to some errors in the process of data extraction and/or image restoration. In order to truly achieve the fully reversible recovery of the original image, *Ma et al.* [15] proposed a novel RDHEI method by reserving the room before encryption. They divide the original image into two parts, A and B, and then embed two or more LSB planes of A into B by employing traditional RDH algorithms. Next, they encrypt the preprocessed image to generate an encrypted image. Thus, locations of these vacated LSBs in the encrypted image can be used to embed information. In [16], *Zhang et al.* proposed an RDHEI method based on an estimation technique. They estimate a small portion of the pixels through a large portion of pixels in the original image and then encrypt the original image using standard encryption algorithms. The final encrypted image is obtained by encrypting the estimating errors and connecting it to the large group of encrypted pixels. Finally, the data-hider can embed additional information by modifying the estimating errors.

*Zhang et al.* [17], encrypted the original plaintext image by using the public key known to the receiver. The data-hider can embed some additional information into the encrypted image by multi-layer wet paper coding without knowing the original image. In the decoding stage, accurate extraction of the embedded data and lossless recovery of the original image can be achieved according to the key. In *Xu and Wang* [18] proposed method, a stream cipher is utilized to encrypt sample pixels and a specific encryption mode is designed to encrypt interpolation-error of non-sample pixels. Then the data-hider can embed secret data into interpolation-error by histogram shifting and difference expansion technique. *Huang et al.* [19] proposed a new simple yet effective framework for RDH in the encrypted domain. In this framework, the original image is encrypted by block permutation and pixel bit-level XOR operation to generate an encrypted image. Then embed the secret information in the encrypted image using the previously proposed RDH methods, such as histogram shifting and difference expansion.

Different from the methods described above, several recent RDHEI methods vacate the room in which additional information is embedded by computing a label map of the original image [20]-[22]. *Puteaux et al.* [20] proposed an MSB predictive detection method to generate a label map of the original image. Then embedding the label map into the encrypted image obtained by bitwise XOR encryption. Finally, the data-hider can embed the secret information according to the label of each pixel in the encrypted image. With a similar idea, an improved RDHEI method proposed by *Puyang et al.* [21]. They use two MSBs to mark each pixel value in the original image, resulting in a significant increase in embedding capacity. In the method of *Yi et al.* [13], they use a small number of pixels as reference values to calculate the prediction error of most pixels and propose a parametric binary tree labeling method to distinguish all prediction errors. In the data hiding stage, the data-hider can embed the secret information according to different prediction errors in the encrypted image with the tag value.

In this paper, we propose a new high-capacity and separable RDHEI method using Huffman coding labeling (HVLCL-RDHEI). We can divide the proposed method into three phases, namely the encryption phase, the embedded phase, and the decryption phase. In the encryption phase, we first calculate the label map of the original image, then encrypt the original image and embed the label map into the encrypted image. In the embedding phase, we can embed multi-bit information in each encrypted pixel by multi-MSB substitution based on the embedded label map. Finally, in the decryption phase, data extraction and image recovery can be performed separately, and the extracted data and the restored image are both lossless. Compared with the previous methods, our method has greatly improved the embedding capacity.

The rest of the paper is organized as follows. Section II describes in detail the latest three related work. Section III mainly introduces the method we proposed. Experimental results are given in Section IV. Section V concludes this paper and explains the future work.

## II. Related Works

In the plaintext domain, most RDH algorithms embed data by modifying the LSB (least significant bit) to ensure the visual quality of the image. But in the encryption domain, we do not require high visual quality of the image. For this reason, the MSB (most significant bit) of the pixel can be used as marked bits to record the label map for image restoration. *Puteaux et al.* [20] proposed a high-capacity reversible data hiding approach with embedded prediction errors (EPE-HCRDH). In order to further improve the embedding ability, *Puyang et al.* [21] proposed a reversible data hiding algorithm in encrypted images with two-MSB labeling (Two-MSB-RDHEI). *Yi et al.* [22] proposed an RDHEI method using parametric binary tree labeling scheme (PBTL-RDHEI).

### A. EPE-HCRDH

*Puteaux et al.* [20] proposed an EPE-HCRDH approach, which aims to exactly reconstruct the original image while keeping high embedding capacity. The proposed method consists of four steps, namely MSB prediction error detection, image encryption and pixel marking, data hiding by MSB substitution, data extraction and image recovery.

1) *MSB Prediction Error Detection*: In this method, the original MSB values are lost in the subsequent data hiding step. So, the first step is to use the previous pixels to predict the current pixel value and mark the wrong pixel. For the current pixel $x(i,j)$ in the original image $I$ with $m \times n$ pixels, which $1 < i \leq m$ and $1 < j \leq n$. Its inverse value is expressed as $inv(i,j) = (x(i,j) + 128) \mod 256$. Compute the predictive value $px(i,j)$ by left and top pixels of $x(i,j)$,

$$px(i, j) = floor(\frac{x(i-1, j) + x(i, j-1)}{2}) \quad (1)$$

Then calculate the absolute difference between $px(i,j)$ and $x(i,j)$ and between $px(i,j)$ and $inv(i,j)$, and record them as $\Delta$ and $\Delta^{inv}$, so that:

$$\begin{cases} \Delta = |px(i,j) - x(i,j)| \\ \Delta^{inv} = |px(i,j) - inv(i,j)| \end{cases} \quad (2)$$

Compare the values of $\Delta$ and $\Delta^{inv}$. If $\Delta < \Delta^{inv}$, there is no prediction error because the original value of $x(i,j)$ is closer to its predictor than the inverse value, and the map of $x(i,j)$ is marked as "0", indicating it can embed data in the encrypted image. Otherwise, there is a prediction error and the map is marked as "1", indicating that the information cannot be embedded.

2) *Image Encryption and Pixel Marking*: In this phase, the original image $I$ is first encrypted and the label map is embedded in the encrypted image. First, they use the encryption key $K_e$ to generate a sequence of pseudo-random bytes $s(i,j)$. Then the encrypted image $I_e$ can then be obtained by exclusive-or (XOR) operation of $x(i,j)$ and $s(i,j)$. Finally, the label map is embedded into the encrypted image $I_e$ by means of MSB replacement to obtain the final encrypted image $I_e$' containing the tag information. Note that there should be done in blocks of eight pixels when embedding the label map, which aims to reduce the error rate during data extraction and image recovery. In addition, if there is at least one error pixel in a block, the MSB of each pixel in the front and the back block of the current block are replaced with "1" as a flag of the current erroneous block.

3) *Data Hiding by MSB Substitution*: Before embedding data, the to-be-inserted message is first encrypted by using the data hiding key $K_w$. Then the MSB of each pixel in the error-free block can be substituted by one-bit secret message $b_k$, with $0 \leq k \leq m \times n$, as shown in Eq. (3). In this way, the marked encrypted image $I_{ew}$ containing secret information can be obtained.

$$x_{ew}(i,j) = b_k \times 128 + (x_e^{'}(i,j) \bmod 128), \quad b_k \in \{0,1\}, \quad 0 \leq k \leq m \times n \quad (3)$$

4) *Data Extraction and Image Recovery*: On the receiving side, different results are obtained depending on the key that is owned. If the receiver only has data hiding key $K_w$, the encrypted secret message can be obtained by extracting the MSB of each pixel in the correct block of the marked encrypted image $I_{ew}$,

$$b_k = x_{ew}(i,j)/128, \quad 0 \leq k \leq m \times n \quad (4)$$

Then, the corresponding original message can be obtained by using the data hiding key $K_w$. If the receiver only has encryption key $K_e$, directly decrypting the marked encrypted image $I_{ew}$ can get the reconstructed image $I'$, which its seven LSBs are same as the original image. The MSB of each pixel can be recovered by prediction error detection method. So, only the receiver has both keys of $K_w$ and $K_e$, data extraction and image recovery can be achieved at the same time.

**B. Two-MSB-RDHEI**

Based on the *Puteaux et al.* [20] algorithm, *Puyang et al.* [21] used two MSB to mark pixels, which greatly improved the embedding capacity. In the Two-MSB-RDHEI method, there are five parts: prediction error detection, image encryption, preprocessing, data hiding, data extraction and image recovery.

1) *Prediction Error Detection*: For the original pixel $x(i,j)$ in the original image $I$ with $m \times n$ pixels, which $1 < i \leq m$ and $1 < j \leq n$, the median edge detection (MED) predictor [7] is used to generate predicted value $px(i,j)$ based on its three neighboring pixels. Calculate the two-MSB values of the current pixel $x(i,j)$ and corresponding predicted value $px(i,j)$:

$$\begin{cases} x^{2MSB} = x(i,j) - (x(i,j) \bmod 64) \\ px^{2MSB} = px(i,j) - (px(i,j) \bmod 64) \end{cases} \quad (5)$$

Then compare the values of $x^{2MSB}$ and $px^{2MSB}$. If $x^{2MSB}$ is equal to $px^{2MSB}$, there is no prediction error of the current pixel $x(i,j)$ and its label is "0". In other words, if $x^{2MSB}$ is not equal to $px^{2MSB}$, the current pixel $x(i,j)$ is an error pixel and it is labeled with "1".

2) *Image Encryption*: In this subsection, the original image $I$ is encrypted by bitwise XOR operation with a pseudo-random sequence $s(i,j)$ and obtained the encrypted image $I_e$. Same as the previous algorithm, the sequence of $s(i,j)$ is generated by the encryption key $K_e$.

3) *Preprocessing*: *prediction error highlighting*: For the encrypted image $I_e$, divide four pixels into one block. According to the label map generated in the process of prediction error detection, if there is no error pixel in a block, the MSB and the second MSB are replaced by "0". If there is at least one error pixel in a block, all pixels in this block unchanged. In addition, the MSB and the second MSB in previous and following blocks are set to "1", which aims to highlight prediction error. So, the final encrypted image $I_e^{'}$ with label map is obtained.

4) *Data hiding*: After receiving the final encrypted image $I_e^{'}$, the secret message can be embedded in the correct block by replacing the two MSBs of the pixel,

$$x_{ew}(i,j) = m_1 \times 128 + m_2 \times 64 + (x_e^{'}(i,j) \bmod 64) \quad (6)$$

where $m_1$ and $m_2$ are secret messages. Note that the to-be-inserted message is encrypted by using the data hiding key $K_w$ before the embedding operation.

5) *Data Extraction and Image Recovery*: As with the EPE-HCRDH algorithm, if two MSBs in the preceding and succeeding blocks of a block are not completely marked as "1", the encrypted data can be directly extracted from the first two MSBs of each pixel in the block. Then the original plaintext data can be obtained with the data hiding key $K_w$. If there is the encryption key $K_e$, the recipient can decrypt the marked encrypted image first and then restore the first two MSBs of the current pixel by its predicted value in the correct blocks, while the pixels in the error block are not changed. In this way, the same restored image as the original image is obtained.

**C. PBTL-RDHEI**

*Yi et al.* [22] propose a PBTL-RDHEI method that keeps spatial correlations within small encrypted image blocks and exploiting the spatial redundancy of the encrypted image to embed secret data. The main steps of the method are parametric binary tree labeling (PBTL), image encryption, pixel grouping, pixel labeling using PBTL, data hiding, data extraction and image recovery.

1) *PBTL*: Suppose that all the pixels in an image are divided into two parts, namely G1 and G2. Then set two parameters $\alpha$ and $\beta$, where $1 \leq \alpha, \beta \leq 7$. For the pixels in G2, use $\beta$ bits of all-zero binary code to mark. For G1, classified all pixels into $n_\alpha$ different sub-categories according to $\alpha$ and $\beta$, where $n_\alpha$ is calculated by:

$$n_\alpha = \begin{cases} 2^\alpha - 1, & \alpha \leq \beta \\ (2^\beta - 1) \times 2^{\alpha-\beta}, & otherwise \end{cases} \quad (7)$$

For the pixel in different sub-categories, use different $\alpha$ bits of binary code to mark. Fig.1 is an illustrative example of different $\alpha$ values in the case of $\beta = 3$.

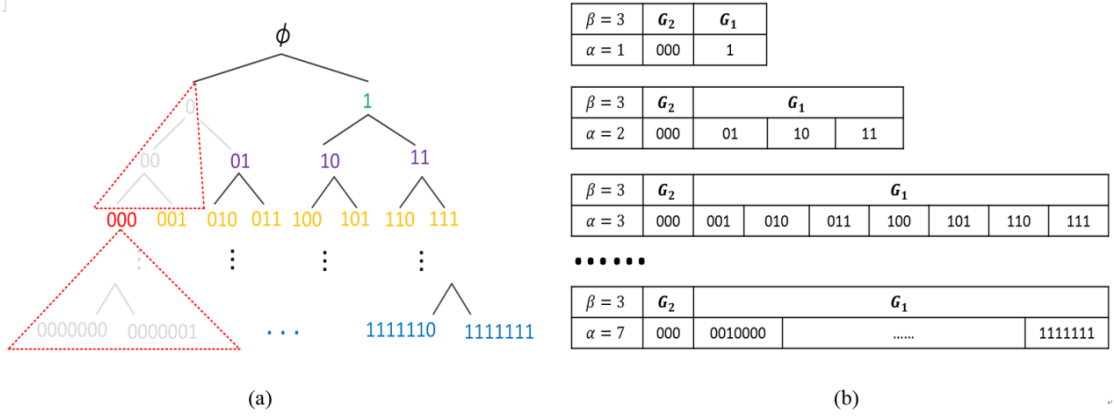

**Fig. 1:** Example of labeling bits selection when β = 3 and α = 1 to 7.

2) *Image Encryption*: Image encryption involves two processes: block permutation and pixel modulation. For a gray-scale image $I$ of size $m \times n$ first divides into $k$ non-overlapped blocks $B_{(i)}$ ($i = 1, 2, ..., k$) with a size of $s \times s$, where $k = mn/s^2$, and $s$ is a small integer that greater than or equal to 2. Then, all blocks are permuted according to encryption key $K_e$ and denoted as $B'_{(i)}$ ($i = 1, 2, ..., k$). For the pixels in the same scrambled block $B'_{(i)}$, the unified modification is performed by:

$$E_{(i)}^j = (B_{(i)}^{'j} + R_i) \bmod 256, \quad (j = 1, 2, ..., s^2) \tag{8}$$

where $B_{(i)}^{'j}$ is the $j^{th}$ pixel of block $B'_{(i)}$ in raster-scan order, $R_i \in [0, 255]$ is a pseudo-random integer generated by encryption key $K_e$.

3) *Pixel Grouping*: After obtaining the encrypted image $E$, divides it into $k$ non-overlapped blocks and separate all pixels into four sets. The first part is the reference pixel ($P_r$), which is used to calculate the prediction error of other pixels in the same block, and it consists of one pixel selected from each block by a user-defined rule. The second part is one special pixel ($P_s$) selected in the first block to store the parameters $\alpha$ and $\beta$. The remaining pixels are divided into two parts, namely embeddable pixel ($P_e$) and non-embeddable pixel ($P_n$), based on their difference value between them and the reference pixel in the same block, where the difference value $e_i$ of current pixel $E_i$ is calculated by:

$$e_i = E_i - E_i^{ref}, \quad i = 1, 2, ..., mn - k - 1 \tag{9}$$

where $E_i^{ref} \in P_r$ is the corresponding reference pixel of $E_i$. If $e_i$ of the current pixel satisfies the following condition, the pixel is divided into $P_e$, conversely, the pixel is divided into $P_n$.

$$\left\lceil -\frac{n_\alpha}{2} \right\rceil \le e_i \le \left\lfloor \frac{n_\alpha - 1}{2} \right\rfloor \tag{10}$$

where $\lceil * \rceil$ and $\lfloor * \rfloor$ are the ceil and floor operations, respectively.

4) *Pixel Labeling using PBTL*: After pixel grouping, the pixels in $P_r$ and $P_s$ remain unchanged, and only the pixels in $P_e$ and $P_n$ are marked by using PBTL method.

5) *Data Hiding*: In the marked encrypted image, the parameters $\alpha$ and $\beta$ are first stored in the special pixel, and the original 8 bits of $P_s$ are stored as auxiliary information in $P_e$. In addition, for the pixels in $P_n$, the $\beta$-bit original bits before labeling need to be recorded as auxiliary information and stored in $P_e$. After pixel labeling, the original value of each pixel in $P_e$ can be obtained with the $\alpha$ bits binary label and the corresponding reference pixel, so that the remaining (8-$\alpha$) bits of each pixel in $P_e$ can embed the information by bit replacement. Of course, for data security, it is necessary to encrypt the secret data to be embedded according to the data hiding key $K_w$ before the embedding operation.

6) *Data Extraction and Image Recovery*: For the receiver, the parameters $α$ and $β$ can be extracted directly. Then group the pixels, the reference pixels $P_r$ and one special pixel $P_s$ are readily available, and the remaining pixels are divided into two parts, $P_e$ and $P_n$, according to the labels of their front $α$ or $β$ bits in the 8-bit binary value. Therefore, the encrypted secret data can be obtained by extracting the remaining $(8-α)$ bits of each pixel in $P_e$, then the original secret data can be obtained by decrypting with the data hiding key $K_w$. On the other hand, in the image restoration process, first, the 8 bits of the special pixel $P_s$ and the first $β$ bits of the pixel in $P_n$ are restored based on the extracted auxiliary information. Then, the image is divided into non-overlapped blocks and the pixels in each block are uniformly decrypted according to the encryption key $K_e$. Finally, the original image is obtained by permuting all blocks inversely.

## III. Proposed Methods

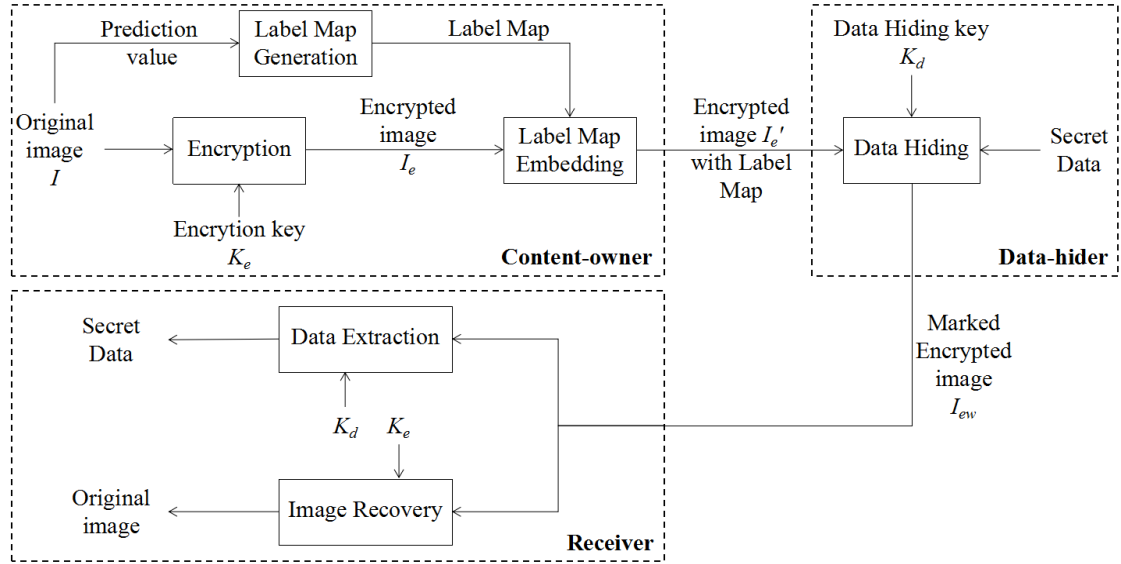

**Fig. 2:** The framework of HVLCL-RDHEI.

The algorithms proposed by *Puteaux et al.* [20] and *Puyang et al.* [21], the 8 bits are used to mark the error block, but there is still a $1/2^8$ error rate in the process of data extraction and image restoration. In the *Yi et al.* [22] proposed algorithm, although it can correctly extract data and losslessly recover images, `the labeling method used is fixed-length coding, which does not make full use of the spatial correlation of pixels in the image. Therefore, based on the correct data extraction and lossless image recovery, in order to increase the embedded capacity as much as possible, we proposed an RDHEI method by multi-MSB substitution using Huffman coding labeling (HVLCL-RDHEI). The framework of the proposed algorithm is shown in Fig. 2. The proposed method consists of three parts, namely content-owner, data-hider, and receiver. First, the content-owner needs to calculate the label map of the original image and encrypt the image, and embed the label map into the encrypted image. Second, the data-hider can extract the label map in the encrypted image and then embed the secret information according to the tag value of each pixel. Finally, the receiver can extract data and recover images based on the key.

## A. Label Map Generation

For the original image $I$ of size $m \times n$, the predicted value of each pixel is first calculated. As shown in Fig. 3, we calculate the predicted value $px(i,j)$ using the MED predictor [7] based on three pixels around the current pixel $x(i,j)$, which $1 < i \leq m$ and $1 < j \leq n$, the formula is as follows:

$$px(i,j) = \begin{cases} \max(x(i-1,j), x(i,j-1)) &, \quad x(i-1,j-1) \leq \min(x(i-1,j), x(i,j-1)) \\ \min(x(i-1,j), x(i,j-1)) &, \quad x(i-1,j-1) \geq \max(x(i-1,j), x(i,j-1)) \\ x(i-1,j) + x(i,j-1) - x(i-1,j-1) &, \quad \text{otherwise} \end{cases} \quad (11)$$

| x(i-1,j-1) | x(i-1,j) |
|---|---|
| x(i,j-1) | x(i,j) |

**Fig. 3:** The context of the current pixel by MED predictor.

Next, we convert the values of $x(i,j)$ and $px(i,j)$ into an 8-bit binary sequence by Eq. (12), denoted as $x^k(i,j)$ and $px^k(i,j)$, where $k=1,2,...,8$.

$$x^k(i,j) = \left\lfloor \frac{x(i,j) \bmod 2^{9-k}}{2^{8-k}} \right\rfloor, \quad k = 1,2,\ldots,8 \quad (12)$$

Then, compare each bit of $x^k(i,j)$ and $px^k(i,j)$ sequentially from first MSB to last LSB until a certain bit is different, and the current pixel's label is equal to its same number of bits. Since the pixel converted binary sequence has 8 bits, the label of the pixel has 9 cases, namely 0 to 8. Assuming that the tag value is represented by $t$, i.e. $t=0,1,2,...,8$, and the maximum value of $t$ obtained according to Eq. (13) is the label of the current pixel $x(i,j)$,

$$\arg\max_{t} \quad x(i,j)^{tMSB} = px(i,j)^{tMSB}, \quad t = 0,1,2,\ldots,8$$
$$\text{subject to} \begin{cases} x(i,j)^{tMSB} = x(i,j) - (x(i,j) \bmod 2^{8-t}) \\ px(i,j)^{tMSB} = px(i,j) - (px(i,j) \bmod 2^{8-t}) \end{cases} \quad (13)$$

where $x(i,j)^{tMSB}$ and $px(i,j)^{tMSB}$ are the $t$-MSB values of $x(i,j)$ and $px(i,j)$, respectively.

After obtained the tag value $t$ of the current pixel, it means that the pixel can embed $(t+1)$ bits in the subsequent data hiding process. The reason is that the front $(t+1)$ MSB of the original pixel can be obtained from its predicted value. In other words, the front $t$-bit MSB of the original pixel is the same as its predicted value, and the $(t+1)^{th}$ MSB can be obtained by negating the value of the corresponding position of its predicted value.

For example, as shown in Fig. 4, assume that the current pixel value $x$ is equal to 156 and its predicted value $px$ is equal to 150. Then $x$ and $px$ are converted into the 8-bit binary sequences, denoted $x^k$ and $px^k$ ($k = 1, 2,…, 8$), respectively, i.e. $x^k = \{1\ 0\ 0\ 1\ 1\ 1\ 0\ 0\}$, $px^k = \{1\ 0\ 0\ 1\ 0\ 1\ 1\ 0\}$. By comparison, it can be obtained that the sequence of $x^k$ and $px^k$ are different in the fifth position, that is, the first four bits are the same, so the label of the pixel $x$ is "$label = 4$", that is, 5 bits of information can be embedded in this pixel.

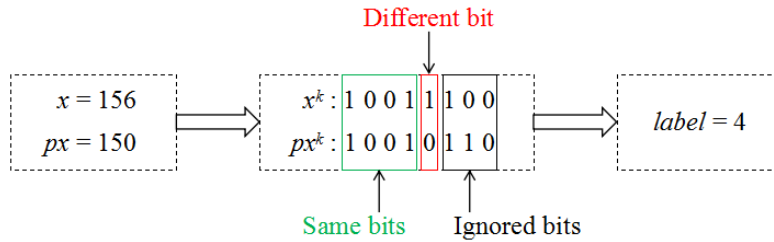

**Fig. 4:** Example of pixel labeling.

Finally, we can scan all the pixels in the image by the above method to obtain the label map of the original image. Note that the pixels in the first row and the first column of the image are reference pixels and are not marked.

**B. Image Encryption**

In this part, each pixel of the original image is encrypted by the encryption key $K_e$. First, we generate a pseudo-random matrix $r(i,j)$ of size $m \times n$ through the key $K_e$. Next, convert the current pixel $x(i,j)$ and its corresponding $r(i,j)$ into the 8-bit binary sequence according to Eq. (12), denoted as $x^k(i,j)$ and $r^k(i,j)$. Then, the following encryption operation is performed,

$$x_e^k(i,j) = x^k(i,j) \oplus r^k(i,j), \quad k=1,2,\ldots,8 \tag{14}$$

where $x_e^k(i,j)$ the encrypted 8-bit binary sequence, and $\oplus$ is the bitwise XOR operation. Finally, the encrypted pixel $x_e(i,j)$ can be calculated by Eq. (15). In this way, we get the encrypted image $I_e$.

$$x_e(i,j) = \sum_{k=1}^{8} x_e^k(i,j) \times 2^{8-k}, \quad k=1,2,\ldots,8 \tag{15}$$

**C. Label Map Coding and Embedding**

According to the label map of the original image obtained from Section III-A, we can calculate the total amount of data that can be embedded in the image. Of course, the label of each pixel needs to be recorded in binary code as auxiliary information and embedded in the encrypted image, the purpose of which is to ensure that the original image can be completely reconstructed. We know that for a natural image, the number of pixels per label is different. Because of this feature, we record the label map through the predefined Huffman coding labeling (HVLCL) rule. The proposed HVLCL method uses variable length coding to effectively compress the amount of auxiliary information, which is equivalent to increasing the embedded payload of the image.

For all the pixels in an image, there are 9 kinds of labels, so we need to use 9 variable length codes to represent each type of label. As shown in Fig. 5, we use 9 kinds of Huffman codings to represent the 9 kinds of labels, namely {00, 01, 100, 101, 1100, 1101, 1110, 11110, 11111}. We first sort the 9 kinds of labels by the number of pixels and then use the shorter code to represent the label with the larger number of pixels. That is, for the 9 variable length codes in Fig. 5, where "*00*" represents the label with the largest number of pixels, and "*11111*" represents the label with the smallest number of pixels.

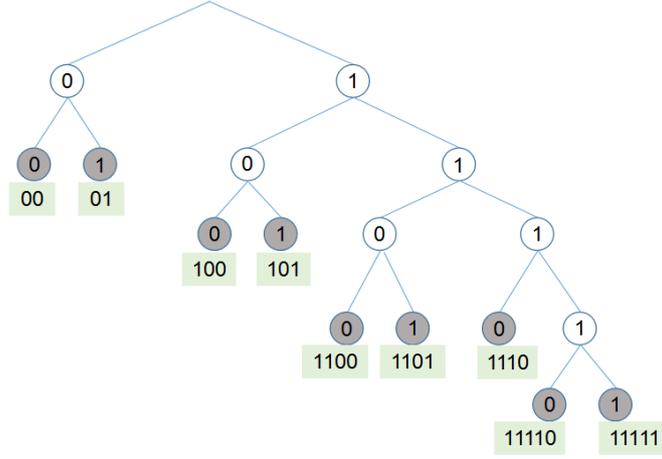

**Fig. 5:** Huffman coding.

For example, in the image of *Lena*, the distribution and the coding of the label map are shown in Table I, where "-1" represents the reference pixel. We can see that the number of pixels with the tag value equal to 5 in *Lena* is the largest, so these pixels are represented by "*00*" code. And the number of pixels with the tag value equal to 1 is the least, so these pixels are represented by "*11111*" code.

**Table I:** *Lena*'s label map distribution and coding.

| Label | -1 | 0 | 1 | 2 | 3 | 4 | 5 | 6 | 7 | 8 |
|---|---|---|---|---|---|---|---|---|---|---|
| Distribution | 1023 | 9818 | 9742 | 15247 | 33246 | 44509 | 53359 | 41758 | 24353 | 29089 |
| Code | - | 11110 | 11111 | 1110 | 101 | 01 | 00 | 100 | 1101 | 1100 |

Now we need to embed the generated label map into the encrypted image before the data hiding operation, in order to generate the space in the encrypted image that can embed the secret data.

In the encrypted image $I_e$, we first convert the label map into a binary sequence through the HVLCL rule. Then, we count the HVLCL rule, the length of the binary sequence, and the binary sequence as auxiliary information. Next, the partial auxiliary information is stored in the reference pixels of the first row and the first column, and the values of these reference pixels are placed behind the auxiliary information. Finally, the remaining auxiliary information and reference pixels are embedded into the encrypted image by multiple MSBs substitution according to the label map, and the embedding formula is as follows:

$$x'_e(i,j) = \begin{cases} x_e(i,j) \bmod 2^{7-t} + \sum_{s=0}^{t}(b_s \times 2^{7-s}), & 0 \le t \le 6 \\ \sum_{s=1}^{8}(b_s \times 2^{8-s}), & 7 \le t \le 8 \end{cases} \quad (16)$$

where *t* is the tag value of the current pixel $x_e(i,j)$ in the $I_e$, and $b_s$ is the auxiliary information in which the current pixel can be embedded. After embedding the auxiliary information and the reference pixels, we get the final encrypted image $I_e'$ containing the label map. Note that in order to be able to completely extract the auxiliary information in subsequent operations, it is necessary to set the plurality of rows and columns as reference pixels in some rough images.

**D. Data Hiding**

In this subsection, we need to extract the auxiliary information from the obtained encrypted image $I_e'$ before the data hiding to recover the label map using HVLCL. First, we extract the partial auxiliary information in the reference pixels of the first row and the first column to obtain the HVLCL mapping rule and the auxiliary information length. Then, according to the existing auxiliary information and HVLCL rule, the tag value $t$ of the current pixel is obtained. Next, the current pixel $x_e'(i,j)$ in the $I_e'$ is converted into an 8-bit binary sequence according to the Eq. (12), and the front $(t+1)$ bits is the auxiliary information embedded in the current pixel. After we have obtained all the auxiliary information, we can restore the label map according to the HVLCL rule. Finally, according to Eq. (16), the secret data is embedded in the remaining pixels, which is the reserved space used to embed the data in the encrypted image. Thus, the marked encrypted image $I_{ew}$ containing secret data is generated.

Note that in order to ensure that the secret data is not extracted directly, we need to encrypt the data by the data hiding key $K_w$ before embedding the secret data.

**E. Data Extraction and Image Recovery**

On the receiving side, first, as with the data hiding process, the receiver can extract the label map and reference pixels from the marked encrypted image $I_{ew}$. Then, based on the label map, the encrypted secret data can be extracted in the same way. Finally, put the reference pixels back to the first row and the first column. The above work can be done without the key, but the next process will get different results depending on the different key that the receiver owned.

If the recipient has only the data hiding key $K_w$, the original secret data can be obtained by directly decrypting the extracted encrypted secret data. However, since there is no encryption key, the original image cannot be reconstructed.

If the recipient has only the encryption key $K_e$, the resulting image is decrypted according to the pseudo-random matrix $r(i,j)$ generated by $K_e$, and the process is the same as Eq. (14). So we get the decrypted image $I_{ew}'$, and only the position of the embedded information in each pixel is different from the original pixel. Next, we scan the pixels in the image except the reference pixels from top to bottom and from left to right. The predicted value $px(i,j)$ of the current pixel $x_{ew}'(i,j)$ is calculated using the MED predictor, and then the original pixel $x(i,j)$ can be restored according to the tag value $t$ and $px(i,j)$. This is because the front $t$-bit MSB of $x(i,j)$ is the same as the corresponding $px(i,j)$, and the $(t+1)^{th}$ MSB of $x(i,j)$ can be obtained by negating the $(t+1)^{th}$ MSB of $px(i,j)$. Note that if the tag value is equal to 8, the original pixel is equal to its predicted value. The recovery process is expressed as follows:

$$x(i,j) = \begin{cases} px(i,j)^{tMSB} + b_{t+1} \times 2^{7-t} + x_{ew}'(i,j) \bmod 2^{7-t} &, 0 \le t \le 7 \\ px(i,j) &, t = 8 \end{cases} \quad (17)$$

where $px(i,j)^{tMSB}$ is the $t$-MSB values of the predicted value $px(i,j)$ obtained by Eq. (13), and $b_{t+1}$ is the $(t+1)^{th}$ binary bit value of the original pixel obtained according to the following formula:

$$b_{t+1} = \begin{cases} 0, & px^{t+1}(i,j) = 1 \\ 1, & px^{t+1}(i,j) = 0 \end{cases} \quad (18)$$

where $px^{t+1}(i,j)$ is the $(t+1)^{th}$ binary bit value of the predicted value $px(i,j)$ obtained by Eq. (12). Finally, the original image is obtained by reconstructing the remaining pixels based on the restored pixels.

Therefore, only there are the data hiding key $K_w$ and the encryption key $K_e$ simultaneously, the recipient can reversibly extract the secret data and restore the original image.

## IV. Experimental Results and Discussion

In this section, we present the experimental results of the HVLCL-RDHEI method and compare it to existing related work. The test image mainly includes six commonly used images, as shown in Fig. 6, which are *Lena*, *Baboon*, *Jetplane*, *Man*, *Airplane* and *Tiffany*. We also experimented with three datasets, BOSSBase [23], BOWS-2 [24], and UCID [25]. We use two metrics with PSNR (Peak signal-to-noise ratio) and SSIM (structural similarity) to evaluate our algorithm reversibility. In addition, we also use *bpp* (bits per pixel) to represent the embedded capacity (i.e. embedding rate) of the algorithm.

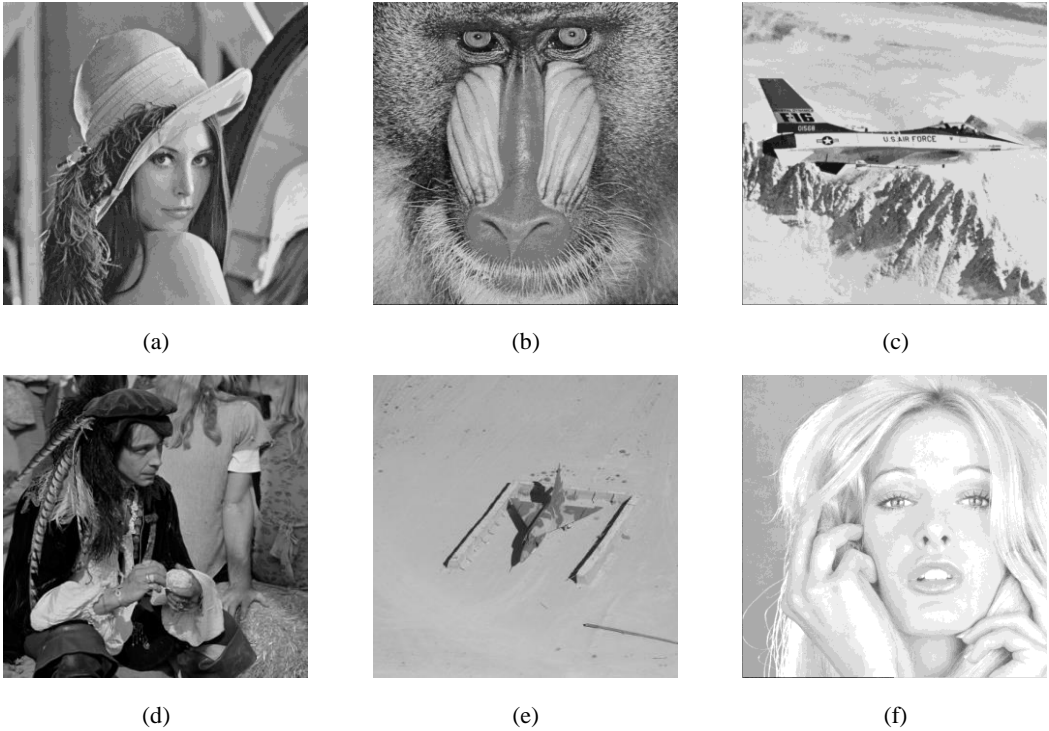

**Fig. 6:** Test image: (a) *Lena*, (b) *Baboon*, (c) *Jetplane*, (d) *Man,* (e) *Airplane*, (f) *Tiffany*.

Section IV-A is the performance analysis of the HVLCL-RDHEI algorithm. Section IV-B shows the results of the proposed method compared with the three algorithms described in Section II, namely *Puteaux et al.*'s EPE-HCRDH approach [20], *Puyang et al.*'s Two-MSB-RDHEI approach [21] and *Yi et al.*'s PBTL-RDHEI approach [22].

### A. Performance Analysis

As we mentioned in Section III-C, the total embedding capacity of an image can be calculated after it's label map has been given. The purpose of using the HVLCL method is to compress the auxiliary information as much as possible to obtain the maximum net payload. Of course, after embedding the data, we can also reversibly extract the data and restore the original image.

Take the image of *Lena* as an example and the main parameters are shown in Table II. The first column is the 9 types of labels in the *Lena* image, where "-1" represents the reference pixel. The second column is the number of pixels for each label, and the third column is the corresponding variable length code obtained by the HVLCL rule. Based on the tag values, we can get the number of data that each pixel can be embedded, so that the total embedding capacity of the *Lena* image is 1470568 bits, as shown in the fourth column of Table II. Similarly, the fifth column is the variable length code length of each pixel, that is, the amount of auxiliary information used to record the label map, for the total of 793304 bits. Obviously, the total capacity minus the amount of auxiliary information gives the net payload equal to 677264 bits. Finally, we subtract the extra 32 bits used to store the HVLCL rule and the extra 20 bits used to store the length of the auxiliary information, and we get the final payload of the *Lena* image, which is 677212 bits.

Similarly, we also calculated the amount of net payload and auxiliary information for other test images, and the results are shown in Table III.

**Table II:** Example of the *Lena*.

| Label | Distribution | Code | Capacity (bits) | Code length (bits) | Payload (bits) |
|---|---|---|---|---|---|
| -1 | 1023 | - | - | - | - |
| 0 | 9818 | 11110 | 1 | 5 | -4 |
| 1 | 9742 | 11111 | 2 | 5 | -3 |
| 2 | 15247 | 1110 | 3 | 4 | -1 |
| 3 | 33246 | 101 | 4 | 3 | 1 |
| 4 | 44509 | 01 | 5 | 2 | 3 |
| 5 | 53359 | 00 | 6 | 2 | 4 |
| 6 | 41758 | 100 | 7 | 3 | 4 |
| 7 | 24353 | 1101 | 8 | 4 | 4 |
| 8 | 29089 | 1100 | 8 | 4 | 4 |
| Total | - | - | 1470568 | 793304 | 677264 |

**Table III:** The embedding capacity and auxiliary information of test images.

| Test Images | Total Capacity (bits) | Code length (bits) | Extra bits (bits) | Payload (bits) |
|---|---|---|---|---|
| Lena | 1470568 | 793304 | 52 | 677212 |
| Baboon | 1074384 | 794941 | 52 | 279391 |
| Jetplane | 1587880 | 793441 | 52 | 794387 |
| Man | 5584907 | 3121508 | 52 | 2463347 |
| Airplane | 1659203 | 682803 | 52 | 976348 |
| Tiffany | 1526934 | 786527 | 52 | 740355 |

After theoretically calculating the payload of the image, we can experiment to analyze the feasibility and reversibility of the proposed HVLCL-RDHEI algorithm. For a 512*512 grey-level images of *Lena*, the original image $I$ is shown in Fig. 7(a), while Fig. 7(b) shows the encrypted image $I_e$ obtained by the encryption key $K_e$. Then, the content owner embeds the label map obtained in Section III-A into the encrypted image to ensure that the data hider can embed the data in the reserved pixel space, and the final encrypted image $I_e'$ containing the label map is shown in Fig. 7(c). It can be seen that the data hider cannot obtain any feature information of the original image from the marked encrypted image, and the security is guaranteed. Fig. 7(d) shows the marked encrypted image $I_{ew}$ after the data hider embeds the secret information according to the label map, and the embedding rate (ER) reaches 2.583 *bpp*. As shown in Fig. 7(e), the receiver can reconstruct the image without error according to the encryption key $K_e$, that is, the PSNR of the reconstructed image is close to 1, and the SSIM is equal to 1. In addition, the recipient can also extract the embedded secret information based on the data key $K_w$.

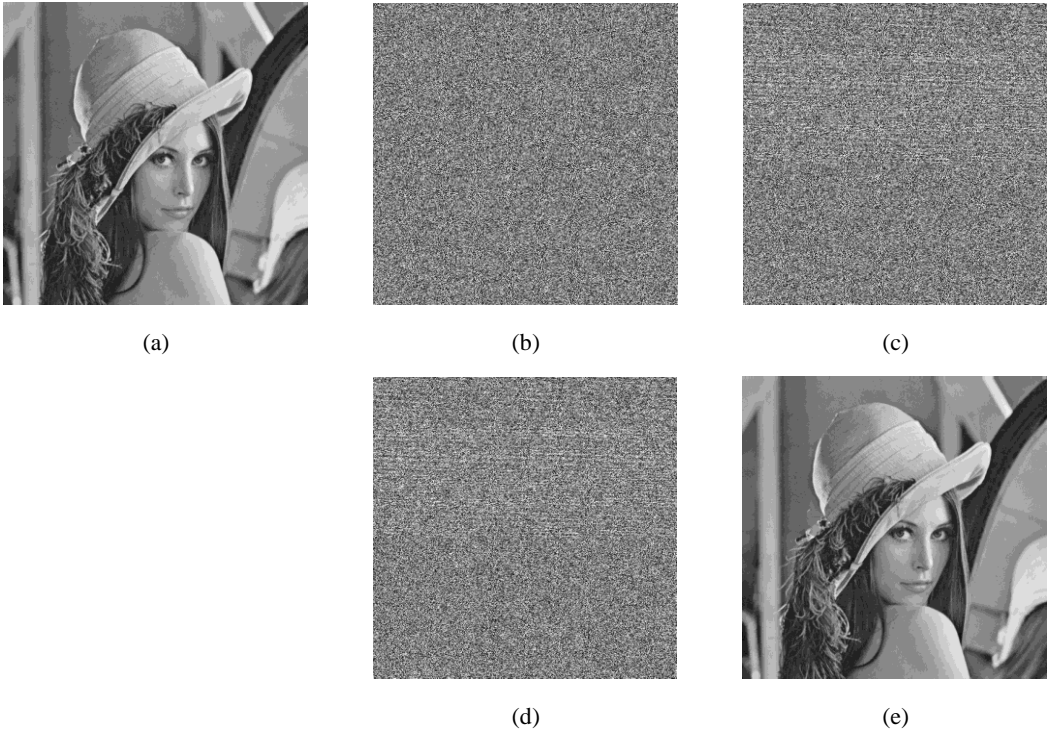

(a)      (b)      (c)

(d)      (e)

**Fig. 7:** Experiment with the proposed HVLCL-RDHEI method, showing the results of each phase: (a) Original image *I*, (b) Encrypted image $I_e$, (c) Marked encrypted image $I_e$' with label map, (d) Loaded encrypted image $I_{ew}$, with the ER = 2.583 *bpp*, (e) Reconstructed image *I*, PSNR $\rightarrow +\infty$, SSIM = 1.

In addition to the test images, we also analyzed the proposed HVLCL-RDHEI algorithm in three datasets, namely BOSSBase [23], BOWS-2 [24] and UCID [25]. Among them, BOSSBase [23] and BOWS-2 [24] have 10,000 grayscale images with a size of 512*512, respectively, while UCID [25] has 1388 grayscale images with a size of 512×384 or 384×512. For relatively smooth images, all pixels have the larger tag value which more information can be embedded, and the encoded auxiliary information is less, so the net payload is large. On the contrary, for rough images, the total embedded amount is small and the auxiliary information is large, so the net payload is small. As shown in Table IV, in the dataset of BOSSBase [23], the image ER reaches 5.898 *bpp* in the best case, and the image ER is only 0.664 *bpp* in the worst case. Similarly, the best case and worst case image ER in the dataset BOWS-2 [24] are 5.622 *bpp* and 0.628 *bpp*, respectively, while the dataset UCID [25] are 5.010 *bpp* and 0.397 *bpp*. In the three datasets, the proposed method has an average ER of 3.361 *bpp*, 3.246 *bpp*, and 2.688 *bpp*. Of course, each image can be extracted without error by the key after embedding the data, and the reconstructed image has a PSNR close to +∞, and SSIM is equal to 1. Therefore, through the above analysis, it is verified that the proposed algorithm has good performance and can be applied to the RDH of the encryption domain.

**Table IV:** Experimental results of three datasets.

| Datasets | Indicators | Best case | Worst case | Average |
|---|---|---|---|---|
| BOSSbase | ER (bpp) | 5.898 | 0.664 | 3.361 |
| | PSNR | +∞ | +∞ | +∞ |
| | SSIM | 1 | 1 | 1 |
| BOWS-2 | ER (bpp) | 5.622 | 0.628 | 3.246 |
| | PSNR | +∞ | +∞ | +∞ |
| | SSIM | 1 | 1 | 1 |
| UCID | ER (bpp) | 5.010 | 0.397 | 2.688 |
| | PSNR | +∞ | +∞ | +∞ |
| | SSIM | 1 | 1 | 1 |

## B. Comparison with State-of-the-arts

In this section, we compare the proposed HVLCL-RDHEI method with several related works. *Puteaux et al.*'s EPE-HCRDH method [20], *Puyang et al.*'s Two-MSB-RDHEI method [21] and *Yi et al.*'s PBTL-RDHEI method [22] both verified the separability of data extraction and image restoration, and the original image can be completely reconstructed by the encryption key. Therefore we will only compare the experimental results of several algorithms by ER (*bpp*).

**Table V:** Comparison of ER (*bpp*) of test images between our method and three state-of-the-art methods.

| Test Image | Puteaux et al.'s EPE-HCRDH [20] | Puyang et al.'s Two-MSB-RDHEI [21] | Yi et al.'s PBTL-RDHEI [22] | Proposed HVLCL-RDHEI |
|---|---|---|---|---|
| Lena | 0.977 | 1.156 | 2.014 | 2.583 |
| Baboon | 0.838 | 0.372 | 0.462 | 1.066 |
| Jetplane | 0.983 | 1.294 | 2.008 | 3.030 |
| Man | 0.981 | 1.152 | 1.7920 | 2.349 |
| Airplane | 0.962 | 1.468 | 2.457 | 3.725 |
| Tiffany | 0.993 | 1.539 | 2.134 | 2.824 |

As shown in Table V, we first compare the experimental results of the proposed HVLCL-RDHEI method with the EPE-HCRDH method, the Two-MSB-RDHEI method and the PBTL-RDHEI method in the six test images. The ER of the EPE-HCRDH method is not more than 1 *bpp*, because the EPE-HCRDH method embeds only one bit of information by MSB replacement in each embeddable pixel. The Two-MSB-RDHEI method is an improved algorithm based on the EPE-HCRDH method, embedding two-bit information at each embeddable pixel, so its ER is improved compared to the EPE-HCRDH method. However, in a rough image of *Baboon*, the ER of the Two-MSB-RDHEI method is lower than the EPE-HCRDH method because there are fewer embeddable pixels and a portion of the embeddable pixels are used to mark the wrong pixels. The PBTL-RDHEI method uses the idea of PBTL to embed multi-bit information in each embeddable pixel, so its ER is high. Like the Two-MSB-RDHEI method, the PBTL-RDHEI method has a lower ER in Baboon, only 0.462 *bpp*. Note that the experimental results of the PBTL-RDHEI method in Table V were obtained with the parameters $α = 5, β = 2$ and the block size of 3×3. The HVLCL-RDHEI method we proposed uses the idea of variable length coding labeling to mark each pixel and then embeds different bits of information by multiple MSB replacement according to the tag value of each pixel. Compared with the three latest methods, the ER of our method has been greatly improved, even in the rough image *Baboon*.

Next, we show the experimental results of the proposed method compared with the three latest methods in the three datasets, namely BOSSBase [23], BOWS-2 [24], and UCID [25]. In the dataset BOSSBase, the average ER of the EPE-HCRDH method is equal to 0.966 *bpp*, while the improved two-MSB-RDHEI method based on the EPE-HCRDH method has an average ER of 1.447 *bpp*. In addition, the average ER of the PBTL-RDHEI method is higher, reaching 1.957 *bpp*. Compared with the three latest algorithms, our proposed HVLCL-RDHEI method has a great improvement in embedding capacity, and the average ER in the dataset BOSSBase reaches 3.361 *bpp*. Similarly, in datasets BOWS-2 and UCID, our proposed algorithm is better than the three latest methods, with an average ER of 3.246 *bpp* and 2.688 *bpp*, respectively, as shown in Table VI.

**Table VI:** Comparison of the average ER (*bpp*) of three datasets between our method and three state-of-the-art methods.

| DataSet | Puteaux et al.'s EPE-HCRDH [20] | Puyang et al.'s Two-MSB-RDHE [21] | Yi et al.'s PBTL-RDHEI [22] | Proposed HVLCL-RDHEI |
|---|---|---|---|---|
| BOSSbase | 0.966 | 1.447 | 1.957 | 3.361 |
| BOWS-2 | 0.968 | 1.346 | 1.881 | 3.246 |
| UCID | 0.893 | 1.179 | 1.586 | 2.688 |

In order to better visualize the experimental results of the proposed HVLCL-RDHEI method, we randomly selected 500 images from three datasets, and obtained the ER of each image by four algorithms, and then the results are shown in Fig. 8. It can be seen that the image ER of our proposed method is generally higher than the previous three latest methods, and only a very small number of images have a low embedding rate. This is due to the fact that the image is not smooth and too much auxiliary information is needed. And for these unsmooth images, the ER obtained by the latest three methods is not very high. In summary, our proposed method is significantly better than the three latest algorithms in terms of embedding capacity.

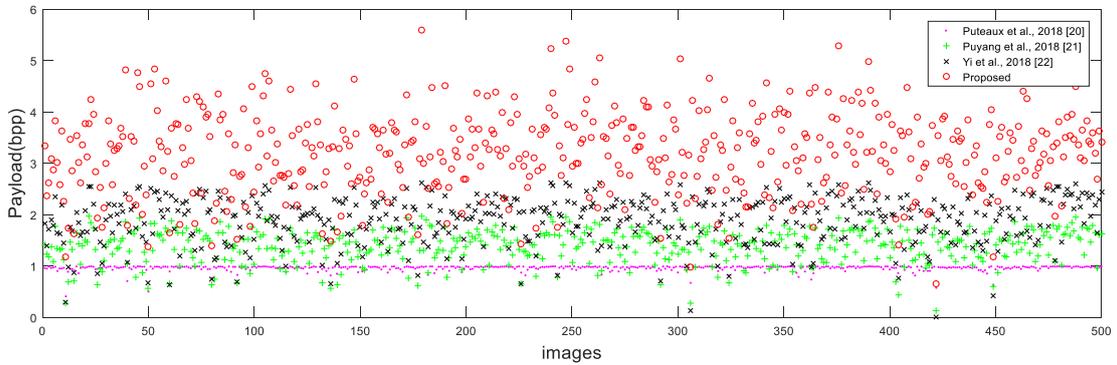

(a)

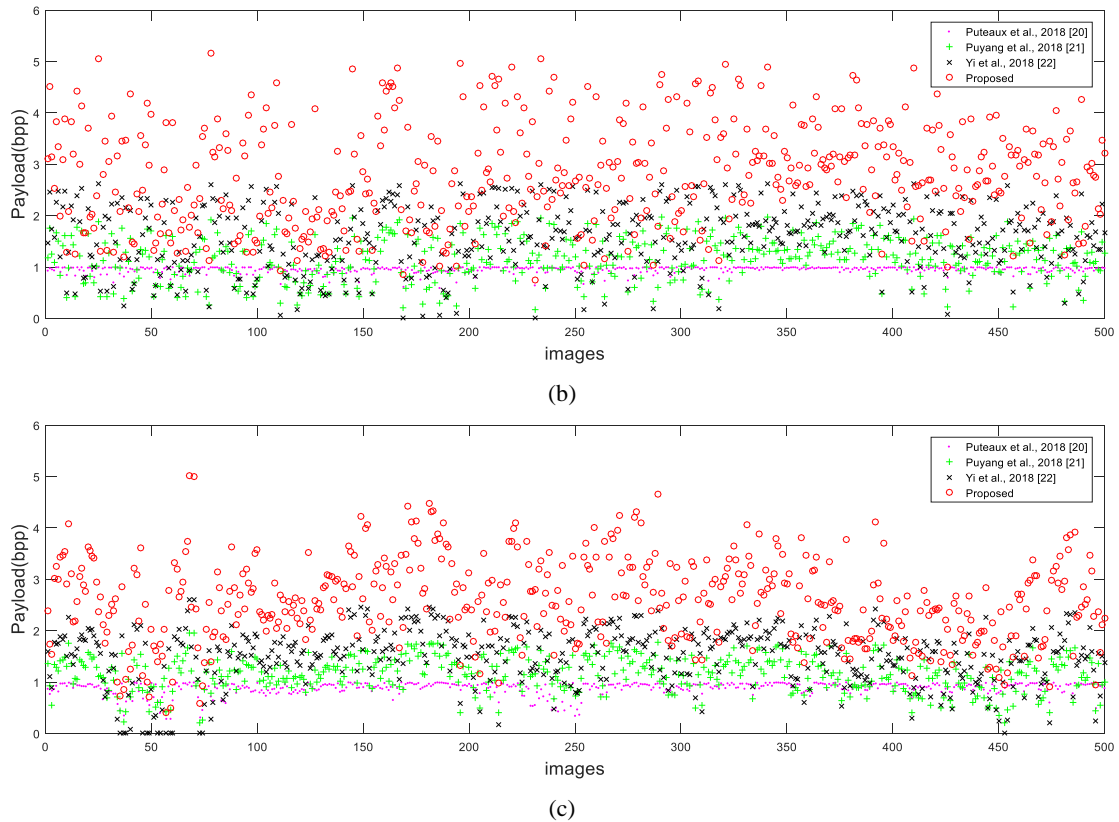

**Fig. 8:** Comparison of ER (*bpp*) between our method and three state-of-the-art methods, which 500 images randomly selected from three datasets. (a) BOSSBase, (b) BOWS-2, (c) UCID.

## V. Conclusion

In this paper, we proposed a Huffman coding labeling (HVLCL) scheme. We first mark the original image with the HVLCL method and embed the label map into the encrypted image. In the process of data hiding, the reserved room can be obtained according to the embedded label map, and then the secret information is embedded in the encrypted image according to the reserved space. During the decryption phase, we can also extract data and recover images based on the embedded label map. The experimental results show that our method has greatly improved the embedding capacity compared to the most advanced algorithms, and at the same time realizes the separability of data extraction and image restoration.

Specifically, in our method, after each pixel finds the tag value based on its predicted value, the total embedded capacity of an original image is determined. The purpose of using the HVLCL method is to compress the label map of the original image to free up more space to embed information. Therefore, in future work, we will focus on the lossless compression processes of an entropy coder, which can further compress the marker information to increase the net embedding capacity.

## Acknowledgments

This research work is partly supported by National Natural Science Foundation of China (61502009, 61872003, U1636206).